\documentclass{elsart}

\usepackage{graphicx}
\usepackage{amsmath}
\usepackage{amssymb}
\newcommand{\rcut}{\ensuremath{r_{\text{cut}}}}

\begin{document}

\begin{frontmatter}

\title{Long-range interactions \& parallel scalability 
       in molecular simulations }

\author[mp]{Michael Patra}
\address[mp]{
Physical Chemistry I, Lund University, Sweden
} 

\author[mh]{Marja T. Hyv\"onen} 
\address[mh]{Wihuri Research Institute, Helsinki, Finland; 
Laboratory of Physics and Helsinki Institute of Physics, 
Helsinki University of Technology, Finland} 

\author[ef]{Emma Falck}
\address[ef]{Beckman Institute for Advanced Science and Technology,
University of Illinois at Urbana-Champaign, Urbana (IL), U.S.A.}

\author[ms]{Mohsen Sabouri-Ghomi}
\address[ms]{Department of Biology,
Virginia Polytechnic Institute \& State University
Blacksburg (VA), U.S.A.; Department of Cell Biology, 
The Scripps Research Institute,
La Jolla (CA), U.S.A.}

\author[iv]{Ilpo Vattulainen}
\address[iv]{Institute of Physics, 
             Tampere University of Technology, Tampere, Finland; 
             Laboratory of Physics and Helsinki Institute of Physics, 
             Helsinki University of Technology, Finland; 
             University of Southern Denmark, Odense, Denmark}

\author[mk]{Mikko Karttunen} 
\address[mk]{Department of Applied Mathematics, the University of 
Western Ontario, London, Ontario, Canada.
}

\begin{abstract}

Typical biomolecular systems such as cellular membranes, 
DNA, and protein complexes are highly charged. Thus, efficient 
and accurate treatment of electrostatic interactions is of 
great importance in computational modelling of such systems.
We have employed the GROMACS   simulation 
package to perform
extensive benchmarking of different 
commonly used electrostatic schemes on a range of computer 
architectures (Pentium-4, IBM Power 4, and Apple/IBM G5)
for single processor and parallel performance up to 8 nodes
-- we have also tested the scalability on four different networks, 
namely Infiniband, GigaBit Ethernet, Fast Ethernet,
and nearly uniform memory architecture, i.e., communication between CPUs is
possible by directly reading from or writing to other CPUs' local memory. 
It turns 
out that the particle-mesh Ewald method (PME) performs 
surprisingly well and offers competitive performance unless 
parallel runs on PC hardware with older network infrastructure are 
needed. Lipid bilayers of sizes 128, 512 and 2048 lipid molecules
were used as the test systems representing 
typical cases encountered in biomolecular simulations. Our results 
enable an accurate prediction of computational speed on most 
current computing systems, both for serial and parallel runs. 
These results should be helpful in, for example, choosing the 
most suitable configuration for a small departmental computer cluster. 
%For serial runs, the speed of PME is comparable to truncation
%schemes. This remains true for parallel runs on high-level computer equipment
%while for standard equipment reaction field becomes advisable.

\end{abstract}

\end{frontmatter} 
\section{Introduction}

During the last few decades, molecular modelling has emerged as an
indispensable tool in studies of soft-matter and
biological systems providing information not accessible by current experimental
techniques~\cite{Schlick:02jq,Scott:02ug,Schlick:99eo,MerzJr.:96qh,Saiz:02dx,Chiu:02tx}. 
One of the challenges in modelling biomolecular systems is the fact that
their properties are largely dictated by electrostatics. At the same time,
biosystems are large -- especially when compared to the size of unit cells of
crystals or typical inorganic molecules -- and characterised by an abundance of
water. The numerical treatment of electrostatics thus needs to be both
accurate and fast.

Electrostatic schemes can be roughly divided into \emph{truncation methods} where all
explicit interactions beyond a certain cutoff are ignored, and  \emph{long-range
methods} which explicitly take all interactions into account. 
There is
one important difference between the two classes: all \emph{long-range} algorithms
give, in principle, the same result and are thus
interchangeable~\cite{Sagui:99eq}. Their error is of numerical nature and can
be tuned by, e.g., adjusting the number of terms included in  series
expansions~\cite{Kolafa:92cl,Deserno:98qj,Deserno:98ec,Greengard:97vg}. 
Selecting one of them for a given application is primarily a
technical decision. This is not the case for \emph{truncation methods}, and the
different truncation schemes are indeed different. Shifting
functions and reaction field techniques are examples of methods
that can be used instead of plain, abrupt truncation. 

There have been many studies comparing different
electrostatic schemes for systems such as water, ions, peptides, proteins, or
lipids. To the best of our knowledge, the first of such articles was published
more than 15 years ago~\cite{Alper:89gj}. Common to almost all of the earlier studies
is the use of a very short cutoff $\rcut\approx 1~\mathrm{nm}$ for truncation. 
Hence, the poor performance of truncation schemes in those studies is
no surprise. Such short cutoffs are neither needed nor should be used 
with present-day computational
resources, and there have been some recent studies on the
differences between electrostatic schemes for cutoffs of at least
$1.4~\mathrm{nm}$, namely for water~\cite{Hess:02kr,Alper:93ra,Alper:93vs},
peptides~\cite{Schreiber:92ar},
proteins~\cite{Faraldo-Gomez:02wg,Tieleman:02tl,York:93mp}, 
and lipids~\cite{Patra:03vl,Anezo:03mz,Patra:04te}.

In this paper, we benchmark some of the most commonly used 
electrostatic schemes through extensive atomic-level MD simulations of
lipid bilayer systems on some of the most common hardware, namely
Pentium-4, IBM Power 4, and Apple/IBM G5. Additionally, we tested the parallel 
performance on Infiniband, GigaBit Ethernet, Fast Ethernet, and
(N)UMA (nearly uniform memory architecture) networks.
GROMACS software was used for the simulations since
it is one of the most popular simulation engines in biophysical and 
soft matter simulations, and comes with open source code. 
There exists other well-established and widely used simulation packages,
such as NAMD~\cite{Phillips:05uq}, Amber~\cite{Case:05mc}, and Charmm~\cite{Brooks:83sz}, 
to name some of the most common ones. 
%%%
%%% new text here
%%%%
While possibly being the least versatile of these, 
GROMACS is generally accepted as the
fasted one on a single processor CPU due to its use of inner loops written in assembler. The increased
speed within each node puts additional stress onto the network in a parallel
simulation, thus allowing a good presentation of parallelisation limitations. 
It is expected that also in the future the speed of processor nodes will keep on increasing faster than the
speed of the networks, and thus we believe that the results 
from Gromacs provide a fairly good
representation of the behaviour on future computers.
%The need for a large number of test runs forced us
%to restrict ourselves to one program.
%%%
%%%
Although it would have been interesting,
we did not have the resources to test other packages than GROMACS -- the
simulations presented here took approximately 22\,000 CPU hours.
General trends should be very similar and independent of the program package.
We would like to point out, however, that parallel performance varies due 
to different implementations
particularly in the case of the particle-mesh Ewald method (PME) as it depends on 
the evaluation of Fourier series -- but again, the general trends are expected
to be the same; many of the common simulation packages use the FFTW library~~\cite{Frigo:05st} for Fast Fourier Transforms. 
Physical aspects, such as the appearance of various artifacts in physical
observables, have been 
addressed in various other articles, e.g. Refs.~\cite{Patra:03vl,Patra:04te},
and will not be discussed here.

This paper is organised as follows. We first describe the modelling of the
bilayer and the different electrostatic
schemes employed here (Secs.~\ref{secModel} and~\ref{secElektro}). 
We then present the results of benchmarking (Sec.~\ref{secResults}). 
The
detailed presentation of the results is designed to allow
the reader to compute how long a simulation of his\,/\,her own will take
using a particular architecture --  or whether such a simulation is
feasible. 
Although we focus on benchmarking, a short discussion of the physical 
aspects is provided in  Sec.~\ref{secResults}. The discussion covers
also more general aspects related to simulations of systems with
Coulomb interactions.

%%%%%%%%%%%%%%%%%%%%%%%%%%%%%%%%%%%%%%%%%%%%%%%%%%%%%%%%%%%%%%%%%%%%%%%%%%%

\section{Model and simulation details}
\label{secModel}

\begin{figure}
\includegraphics[width=\columnwidth]{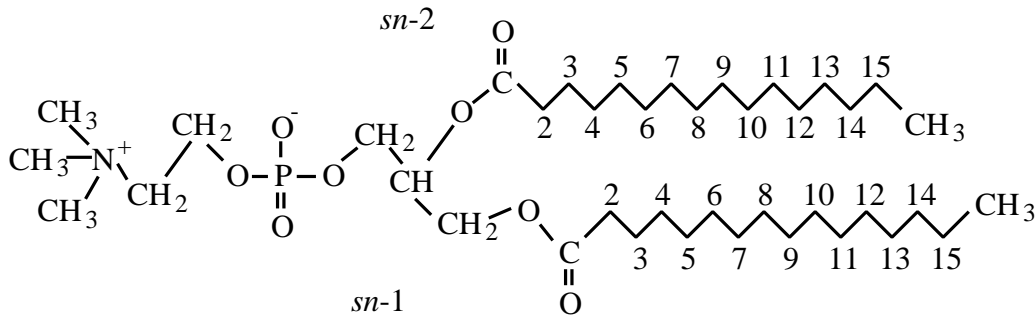}
\caption{DPPC molecule used in all bilayer simulations here.
%
%including the numbering 
%of hydrocarbon tails used in Sec.~\protect\ref{secOrder}.
}  
\label{figDPPC}
\end{figure}

As our benchmark system we used 
a single-component lipid bilayer comprised of 128  dipalmitoyl
phosphatidylcholine (DPPC) molecules, see Fig.~\ref{figDPPC}, fully hydrated by
3655 water molecules. 
To see the effect of increasing system size, membranes consisting of
512 and 2048 DPPC molecules were also simulated.
DPPC molecules are described by the model from
Ref.~\cite{Tieleman:96lr}, available at
\\
http://moose.bio.ucalgary.ca/Downloads/files/dppc.itp, which utilises
the description of lipids from Ref.~\cite{Berger:97el}, available at 
\\
http://moose.bio.ucalgary.ca/Downloads/files/lipid.itp. 
The rigid SPC water model~\cite{Berendsen:81px} was used for water. 
The simulations were  started
from a structure published in Ref.~\cite{Tieleman:96lr} (run E), available
at 
\\
http://moose.bio.ucalgary.ca/Downloads/files/dppc128.pdb.

The simulations were performed using the GROMACS simulation package~\cite{Lindahl:01fx}.
The bond lengths of DPPC molecules were constrained by the LINCS
algorithm~\cite{Hess:97zl} while SETTLE~\cite{Miyamoto:92up} was used for water. 
Those choices allowed a time step of
$2.0~\mathrm{fs}$. For computing nonbonded interactions, we employed
the twin-range setup~\cite{Bishop:97fm}, i.e., interactions within
$r_{\mathrm{list}}=1.0~\mathrm{nm}$ 
%(called ``short-range part'') are
were evaluated at every integration step, and interactions outside of 
$r_{\mathrm{list}}$ only every tenth integration time step.
%{\bf Ilpo's comment: Definition of charge groups?} 

Lennard--Jones interaction was cut off at $1.0~\mathrm{nm}$ (with one
exception, see Sec.~\ref{secReactionField}). DPPC and water molecules were
separately coupled to a heat bath at temperature $T = 323~\mathrm{K}$, and the
pressure was kept at $1~\mathrm{bar}$, both using the Berendsen
algorithms~\cite{Berendsen:84qu}. The size of the simulation box in the plane of
the bilayer ($x$-$y$ plane) was allowed to fluctuate independently of its
height.
We present results of a few hundred short simulations with different
system sizes and computer \& network architectures. 
This was supplemented with several longer simulations,
to compute essential quantities needed to characterise the bilayers,
such as the area per lipid and, and to determine how these quantities depend on
the chosen electrostatics scheme.
%
%Initially, 
Equilibrium was monitored the commonly used
way by measuring the
area per lipid.
%~\cite{Patra:03vl,Patra:04te}.
In total, the simulations took about 22\,000 hours of CPU time.
%That the systems had reached equilibrium was determined 
%
%We present 11 different simulations of $50~\mathrm{ns}$ each,
%representing a total of 22\,000 hours of CPU time. From monitoring the area per
%lipid we found that after typically $20~\mathrm{ns}$ the systems had reached
%equilibrium. We thus discarded the first $20~\mathrm{ns}$ and used the
%remaining $30~\mathrm{ns}$ for analysis.

%%%%%%%%%%%%%%%%%%%%%%%%%%%%%%%%%%%%%%%%%%%%%%%%%%%%%%%%%%%%%%%%%%%%%%%%%%%%%%

\section{Schemes for electrostatic interactions}
\label{secElektro}

The electrostatic interaction between two point charges $q_i$ and $q_j$ 
depends only
on their mutual separation $r$ and is given by the well-known Coulomb law
\begin{equation}
	V_{ij}(r) = \frac{q_i q_j}{4\pi \epsilon_0 r} \;,
	\label{eqPotential1}
\end{equation}
and the total potential then follows from
\begin{equation}
	V_{\text{total}} = \sum_{i<j} V_{ij}(r_{ij})\;.
	\label{eqPotential2}
\end{equation}
As this interaction is long-ranged,
the potential has to be replaced by some related potential 
$\mathcal{V}(r)$ that offers the advantage of being easier to compute.

\subsection{Truncation methods} 

Truncation methods set $\mathcal{V}_{ij}(r)=0$ for $r$ larger than some cutoff
\rcut. This reduces the computational cost from $\mathcal{O}(N^2)$ -- all
combinations of the $N$ particles inside the simulation volume have to be
considered -- to $\mathcal{O}(N r_{\text{cut}}^3)$. The saving is substantial
if $r_{\text{cut}}$ is much smaller than the extension of the simulation box.
For the present study, we used $\rcut=1.8~\mathrm{nm}$,
$\rcut=2.0~\mathrm{nm}$, and $\rcut=2.5~\mathrm{nm}$. Truncation can be done in 
different ways, the usual approaches being abrupt truncation, using shifting functions,
or using the reaction field method.

\subsubsection{Abrupt truncation}

Conceptually the easiest approach is to set $\mathcal{V}_{ij}(r)=V_{ij}(r)$ for
$r\le r_{\text{cut}}$, and $\mathcal{V}_{ij}(r)=0$ otherwise. 
This scheme leads to discontinuities in both electrostatic potential and
force at the cutoff. The simulations done using this scheme are labelled by
their value of $r_{\text{cut}}$ throughout this paper.

\subsubsection{Shifting functions}

The abrupt truncation of the potential at $r_{\text{cut}}$ can be avoided by
modifying the potential $\mathcal{V}_{ij}(r)$ for $r_{\text{switch}}\le r  \le
\rcut$ such that both the resulting force and its derivative are continuous for all
$r$. This is easily
achieved by replacing $\mathcal{V}_{ij}(r)$ in this interval by a third-order
polynomial. Its coefficients are uniquely defined by the continuity
requirements. Note that neither the potential nor the force are simply
``shifted''. The name ``switching function'' thus captures the concept
better than the commonly used name ``shifting function''. 
%Unfortunately, the latter term seems to be used more often.

The choice for $r_{\text{switch}}$ is, in principle, arbitrary. The two most
common choices are $r_{\text{switch}}=0$, or to set $r_{\text{switch}}=r_{\mathrm{list}}$,
i.e., the range of short-range interactions. We adopt the
latter and thus set $r_{\text{switch}}=1.0~\mathrm{nm}$. We label the
simulations employing a shifting function by their value of $\rcut$ with the
acronym SH.

\subsubsection{Reaction field technique}
\label{secReactionField}

The Onsager reaction field technique~\cite{Onsager:36ix} handles all electrostatic
interactions explicitly within the cutoff distance $r_{\text{cut}}$. For
$r>r_{\text{cut}}$
the system is treated on a mean-field level and is thus completely
described by its dielectric constant
$\epsilon$. The
potential is
%\begin{multline}
\begin{equation}
	\mathcal{V}(r)=\frac{q_i q_j}{4\pi\epsilon_0 r}
	\left[ 1 + \frac{\epsilon -1}{2\epsilon+1} 
	\left(\frac{r}{r_{\text{cut}}}\right)^3 \right] \\
	- \frac{q_i q_j}{4\pi\epsilon_0 r_{\text{cut}}}
	\frac{3\epsilon}{2 \epsilon+1}
	\quad\text{for}\quad r\le r_{\text{cut}}\;.
	\label{eqReactionField}
\end{equation}
%\end{multline}
The second term makes the potential vanish at $r=r_{\text{cut}}$. 

%We want to
%note that the total potential can no longer be written as a sum of pair
%potentials as the interaction of each point charge with the dielectric results
%in an additional energy term. Since this term does
%not depend on particle positions, it is in practise neglected.

A reaction field description needs an additional input parameter, namely the
dielectric constant $\epsilon$. Its choice may be problematic in inhomogeneous
systems such as hydrated lipid membranes, where  $\epsilon \approx 80$ for
water, $\epsilon \approx 25$ at the water-membrane interface and $\epsilon
\approx 4$ inside the lipid bilayer~\cite{Koehorst:04vg}. 
We have used $\epsilon=80$.

%\textbf{SPC water actually needs 60...}

%In
%addition to position, the effective dielectric constant for membrane systems
%also depends on direction. The interactions of the headgroup atoms, for example, 
%can be described by $\epsilon \approx 80$ for in-plane interaction and
%$\epsilon \approx 10$ for perpendicular interaction.

We mark reaction field results by the acronym RF. Since
reaction field has been suggested to allow a smaller $\rcut$ than other
truncation schemes due to its mean-field accounting of the ``cut-away'' medium,
we have included a simulation with a very short cutoff $\rcut=1.2~\mathrm{nm}$.
In order to keep a twin-range setup, we also had to use a smaller
$r_{\mathrm{list}}=r_{\mathrm{vdW}}=0.9~\mathrm{nm}$. (All other simulations
were run using our standard choice
$r_{\mathrm{list}}=r_{\mathrm{vdW}}=1.0~\mathrm{nm}$.)

\subsection{Long-range methods}

It is possible to have $\mathcal{V}_{ij}(r) \ne 0$ for an arbitrarily large $r$.
Due to the finite size of the simulation box this implies
that periodic boundary conditions are built as an integral
part into the electrostatics evaluation scheme. Applying
periodic boundary conditions and long-range electrostatics
means that not only the direct interaction between
two particles is considered but also
the interaction between \textit{all} of their mirror images.
Considering the simulation box as an entity, what one
really wants is that \textit{i)} all the particles in the mirror image
of the simulation box have the same mutual correlations
as in the original simulation box, and \textit{ii)} all particles in
the mirror image of the simulation box are uncorrelated
with their "`twin"' in the original box. All long-range electrostatics
schemes fulfil property \textit{i)} exactly (in contrast
to RF where this property is only fulfilled on a mean-field
level) whereas they obviously totally fail property \textit{ii)}. 
%Artifacts
%%caused by this nonphysical periodicity have been
%observed [29Ð31], see Ref. 32 for a thorough discussion.

Long-range electrostatics are usually computed either by using fast
multipole methods (FMM)~\cite{Greengard:97vg}, the (smooth) Particle-Mesh Ewald
(PME) method,~\cite{Essman:95tm} or the
particle-particle\,/\,particle-mesh method
(P$^3$M)~\cite{Hockney:88sk,Deserno:98qj}. 
Both P$^3$M and PME are based on 
Ewald summation.
In the Ewald summation the electrostatic potential can be written as~\cite{Allen:1987}
\begin{eqnarray}
U_\mathrm{Ewald}(r) = \frac{1}{2} \sum_{i=1}^N \sum_{j=1}^N \sum_{|{\bf n}|=0}^{\infty} \mathrm{'}
q_iq_j \frac{\mathrm{erfc\,}(\alpha|{\bf r}_{ij}+ {\bf n}|)}{|{\bf r}_{ij}+ {\bf n}|}\\ \nonumber
+\frac{1}{2\pi V} \sum_{i=1}^N\sum_{j=1}^N \sum_{{\bf k} \ne 0}
q_iq_j \frac{4\pi^2}{k^2}\exp \left(-\frac{k^2}{4 \alpha} \right) 
%\\
\times \cos ({\bf k} \cdot {\bf r}_{ij}) - \frac{\alpha}{\sqrt{\pi}}
 \sum_{i=1}^Nq_i^2 
 %+ J({\bf M},P)
\label{eq:ewald}
\end{eqnarray} 
Here $q_i$ and $ q_j$ are the two charges, $\alpha$ is the Ewald parameter, ${\bf k}$ is the reciprocal
lattice vector, and $V =L_x \times L_y \times L_z$ is the volume of the system.
The prime means that the term  $i=j$ should be dropped for ${\bf n}=0$ (the
actual simulation box).
Solving the above gives, at best, $\mathcal{O}(N^{3/2})$ scaling. PME and P$^3$M
are based on formulating the above equation in such a way that one can use the Fast Fourier
Transform to obtain better scaling, $\mathcal{O}(N\ln N)$.

FMM offers the advantage of
better scaling with system size [$\mathcal{O}(N)$ compared to $\mathcal{O}(N
\ln N)$] of PME, but is more sensitive to numerical noise, more laborious 
to implement and is currently not included in major simulation packages. 
At the present time PME is the most
frequently used algorithm, and we also use PME here. 
It should be stressed
again that all methods for long-range electrostatics give the same result up to
a numerical error (depending on grid spacing, number of $k$ vectors, etc.) that
can be made arbitrarily small in a controlled way at the expense of an increase
in computational cost~\cite{Deserno:98qj}.

%%%%%%%%%%%%%%%%%%%%%%%%%%%%%%%%%%%%%%%%%%%%%%%%%%%%%%%%%%%%%%%%%%%%%%%%%%%%

\section{Results -- speed and scalibility}
\label{secResults}

%%\section{Benchmarking -- speed and scalibility}
%%\label{secSpeed}

\begin{table}
{\scriptsize\mbox{}\hspace{-2mm}%
\begin{tabular}{|l|rrrr|rrrr|rrrr|}
\hline
\multicolumn{13}{|c|}{\textbf{Pentium-4 2.4 Ghz (Fast Ethernet)}}\\
\hline
&\multicolumn{4}{c|}{$8\times 8$ lipids}
&\multicolumn{4}{c|}{$16\times 16$ lipids}
&\multicolumn{4}{c|}{$32\times 32$ lipids} \\
nodes & 
\multicolumn{1}{c}{1} & \multicolumn{1}{c}{2} &
         \multicolumn{1}{c}{4} & \multicolumn{1}{c|}{8} & 
\multicolumn{1}{c}{1} & \multicolumn{1}{c}{2} &
         \multicolumn{1}{c}{4} & \multicolumn{1}{c|}{8} & 
\multicolumn{1}{c}{1} & \multicolumn{1}{c}{2} &
         \multicolumn{1}{c}{4} & \multicolumn{1}{c|}{8}
\\
\hline
1.8&   17.4 &    10.1 &     8.9 & 7.1 &    
        77.9 &    47.4 &    33.3 & 30.4 &   
        322.1 &   186.1 &   123.1 &    89.2 \\ 
2.0&   20.4 &    12.6 &     9.2 &     7.2 &    
        89.3 &    52.2 &    35.7 &    30.7 &   
        368.3 &   222.2 &   131.4 &    96.4 \\
2.5&   27.9 &    16.4 &    11.4 &     8.2 &   
        130.3 &    71.9 &    43.9 &    36.5 &   
        543.6 &   307.5 &   172.1 &   118.5 \\
\hline
SH 1.8&   22.8 &    14.6 &    10.3 &     7.8 &   
        101.2 &    58.1 &    38.1 &    31.8 &   
        418.5 &   245.6 &   142.9 &   100.7 \\ 
SH 2.0&   25.8 &    16.2 &    11.0 &     8.1 &   
        114.7 &    63.5 &    41.1 &    34.9 &   
        475.7 &   266.4 &   158.5 &   106.1 \\
SH 2.5&   35.8 &    23.1 &    13.5 &     9.4 &   
        165.7 &    82.8 &    51.2 &    40.1 &   
        672.8 &   359.3 &   208.1 &   138.2 \\
\hline
RF 1.2&   11.5 &     8.5 &     7.4 &     6.2 &    
        49.7 &    36.2 &    26.7 &    26.7 &   
        201.0 &   141.1 &    95.3 &    77.4 \\ 
RF 1.8&   17.1 &    11.4 &     8.9 &     7.1 &    
        82.1 &    49.2 &    33.5 &    30.1 &   
        328.5 &   195.3 &   123.2 &    89.3 \\
RF 2.0&   20.0 &    13.5 &     9.6 &     7.4 &    
        91.9 &    53.9 &    35.8 &    31.7 &   
        380.1 &   221.4 &   135.0 &    96.5 \\
RF 2.5&   28.8 &    17.8 &    11.7 &     8.3 &   
        135.3 &    70.3 &    46.0 &    35.8 &   
        558.5 &   311.5 &   177.6 &   121.4 \\
\hline
PME&   27.4 &    39.4 &    63.6 &    82.2 &   
        116.9 &   154.9 &   244.9 &   323.9 &   
        480.0 &   624.7 &   964.4 &  1178.6 \\
\hline
\multicolumn{13}{c}{ }\\
\hline
\multicolumn{13}{|c|}{\textbf{Pentium-4 2.66 Ghz (Gigabit Ethernet)}}\\
\hline
%&\multicolumn{4}{c|}{$8\times 8$ lipids}
%&\multicolumn{4}{c|}{$16\times 16$ lipids}
%&\multicolumn{4}{c|}{$32\times 32$ lipids} \\
nodes & 
\multicolumn{1}{c}{1} & \multicolumn{1}{c}{2} &
         \multicolumn{1}{c}{4} & \multicolumn{1}{c|}{8} & 
\multicolumn{1}{c}{1} & \multicolumn{1}{c}{2} &
         \multicolumn{1}{c}{4} & \multicolumn{1}{c|}{8} & 
\multicolumn{1}{c}{1} & \multicolumn{1}{c}{2} &
         \multicolumn{1}{c}{4} & \multicolumn{1}{c|}{8}
\\
\hline
1.8&   15.3 &     7.4 &     5.3 &     3.6 & 
        73.1 &    37.9 &    21.8 &    15.7 &   
        295.0 &   147.2 &    81.4 &    52.2 \\
2.0&   17.8 &    10.3 &     6.5 &     3.6 &    
        81.9 &    41.7 &    24.2 &    15.8 &   
        334.0 &   178.8 &    91.4 &    54.6 \\
2.5&   25.3 &    12.6 &     7.8 &     4.9 &   
        118.1 &    54.2 &    31.0 &    21.7 &   
        497.2 &   244.2 &   129.9 &    76.4 \\
\hline
SH 1.8&   20.7 &    11.5 &     7.1 &     4.6 &    
        93.9 &    44.0 &    26.8 &    19.3 &   
        370.3 &   188.6 &   102.6 &    60.8 \\ 
SH 2.0&   23.5 &    12.2 &     7.4 &     4.6 &   
        109.2 &    53.9 &    29.2 &    19.7 &   
        431.2 &   212.1 &   115.8 &    65.8 \\
SH 2.5&   32.4 &    17.9 &    10.1 &     6.1 &   
        152.4 &    68.1 &    38.2 &    23.1 &   
        605.7 &   320.0 &   165.4 &    92.2 \\
\hline
RF 1.2& 10.3 &     5.8 &     3.8 &     2.8 &    
        46.1 &    25.8 &    14.6 &    10.4 &   
        184.6 &    93.9 &    60.1 &    33.2 \\
RF 1.8&   16.1 &     8.6 &     5.0 &     3.8 &    
        72.9 &    36.7 &    16.4 &    15.3 &   
        299.7 &   148.6 &    84.2 &    56.1 \\
RF 2.0&   18.5 &    10.8 &     6.1 &     4.2 &    
        84.0 &    39.9 &    23.9 &    18.3 &   
        347.4 &   181.9 &    89.4 &    56.4 \\
RF 2.5&   25.4 &    13.3 &     7.6 &     5.4 &   
        119.9 &    56.2 &    31.2 &    20.7 &   
        512.8 &   251.8 &   138.1 &    81.5 \\
\hline
PME&   25.0 &    20.3 &    17.6 &    20.8 &   
        108.8 &    82.8 &    81.4 &    79.7 &   
        428.9 &   329.6 &   283.8 &   243.9 \\
\hline
\multicolumn{13}{c}{ }\\
\hline
\end{tabular}}
\caption{Hours of wall clock time needed per $1~\mathrm{ns}$ of trajectory.
The different columns correspond to different membrane sizes and different
numbers of nodes (for parallel runs).
The
specified membrane size is the number of lipids per single leaflet. The studied bilayers 
thus contained
twice that number of lipids.}
\label{tabSpeed1}
\end{table}
%\end{tabular}
%\end{table}

\vspace*{-1cm}
\begin{table}
{\scriptsize\mbox{}\hspace{-2mm}%
\begin{tabular}{|l|rrrr|rrrr|rrrr|}
\hline
\multicolumn{13}{|c|}{\textbf{Power-4 1.1 Ghz ((N)UMA nearly uniform memory access)}}\\
\hline
&\multicolumn{4}{c|}{$8\times 8$ lipids}
&\multicolumn{4}{c|}{$16\times 16$ lipids}
&\multicolumn{4}{c|}{$32\times 32$ lipids} \\
nodes & 
\multicolumn{1}{c}{1} & \multicolumn{1}{c}{2} &
         \multicolumn{1}{c}{4} & \multicolumn{1}{c|}{8} & 
\multicolumn{1}{c}{1} & \multicolumn{1}{c}{2} &
         \multicolumn{1}{c}{4} & \multicolumn{1}{c|}{8} & 
\multicolumn{1}{c}{1} & \multicolumn{1}{c}{2} &
         \multicolumn{1}{c}{4} & \multicolumn{1}{c|}{8}
\\
\hline
1.8&   27.9 &    13.8 &     7.2 &     3.9 &   
        118.5 &    55.6 &    30.0 &    17.2 &   
        471.8 &   232.9 &   130.0 &    71.0 \\
2.0&   32.1 &    15.8 &     8.8 &     5.0 &   
        131.1 &    63.6 &    37.8 &    20.8 &   
        572.1 &   269.4 &   167.4 &    84.6 \\
2.5&   44.9 &    21.8 &    12.2 &     5.7 &   
        191.4 &    87.9 &    50.0 &    25.7 &   
        759.2 &   323.9 &   205.8 &   100.6 \\
\hline
SH 1.8&   49.2 &    23.6 &    11.9 &     6.4 &   
        180.6 &    97.8 &    53.9 &    30.1 &  
        870.8 &   373.2 &   206.5 &    95.1 \\
SH 2.0&   55.6 &    27.2 &    14.9 &     6.5 &   
        212.4 &   109.0 &    58.2 &    34.6 &   
        916.7 &   421.5 &   239.9 &   141.2 \\
SH 2.5&   72.6 &    37.1 &    21.5 &    11.4 &   
        297.8 &   146.2 &    84.7 &    48.1 &  
        1246.7 &   600.4 &   419.3 &   197.5 \\
\hline
RF 1.2&   21.7 &     9.3 &     4.7 &     2.8 &   
        102.8 &    36.8 &    20.8 &    10.6 &   
        256.7 &   153.5 &    85.6 &    48.3 \\
RF 1.8&   31.5 &    15.1 &     7.8 &     4.4 &   
        109.7 &    62.2 &    34.9 &    17.1 &   
        504.4 &   245.6 &   142.2 &    85.3 \\
RF 2.0&   31.9 &    16.8 &     8.8 &     5.6 &   
        122.2 &    68.3 &    36.4 &    19.9 &   
        570.8 &   296.9 &   150.7 &    94.2 \\
RF 2.5&   48.2 &    23.6 &    13.5 &     6.2 &   
        197.4 &    93.6 &    56.8 &    25.6 &   
        819.9 &   437.8 &   221.1 &    67.1 \\
\hline
PME&   42.4 &    21.9 &    11.2 &     7.6 &   
        172.4 &    89.3 &    45.4 &    27.6 &   
        705.0 &   401.9 &   226.1 &   125.7 \\
\hline
\multicolumn{13}{c}{ }\\
\hline
\multicolumn{13}{|c|}{\textbf{G5 2.3 GHz (Infiniband)}}\\
\hline
%&\multicolumn{4}{c|}{$8\times 8$ lipids}
%&\multicolumn{4}{c|}{$16\times 16$ lipids}
%&\multicolumn{4}{c|}{$32\times 32$ lipids} \\
nodes & 
\multicolumn{1}{c}{1} & \multicolumn{1}{c}{2} &
         \multicolumn{1}{c}{4} & \multicolumn{1}{c|}{8} & 
\multicolumn{1}{c}{1} & \multicolumn{1}{c}{2} &
         \multicolumn{1}{c}{4} & \multicolumn{1}{c|}{8} & 
\multicolumn{1}{c}{1} & \multicolumn{1}{c}{2} &
         \multicolumn{1}{c}{4} & \multicolumn{1}{c|}{8}
\\
\hline
1.8 &
	11.0 &     5.4 &     3.1 &     1.9 &    
	52.8 &    22.8 &    11.9 &     7.1 &   
	237.8 &   100.7 &    52.6 &    28.8 \\
2.0 &
	12.8 &     6.1 &     3.6 &     2.2 &    
	60.6 &    25.6 &    13.6 &     8.2 &   
	271.9 &   115.8 &    61.2 &    32.8 \\
2.5 &
	17.9 &     8.6 &     4.6 &     3.1 &   
	 95.3 &    35.6 &    19.0 &    10.7 &   
	 380.7 &   161.2 &    92.8 &    47.8 \\
\hline
SH 1.8&
	16.1 &     7.6 &     4.2 &     2.6 &    
	73.2 &    31.5 &    16.1 &     9.4 &   
	315.7 &   140.4 &    70.1 &    38.1 \\
SH 2.0&
	18.1 &     8.9 &     4.7 &     2.6 &    
	84.9 &    35.6 &    18.6 &    10.6 &   
	360.3 &   153.8 &    80.3 &    42.8 \\
SH 2.5&
	24.6 &    11.9 &     6.4 &     3.8 &   
	127.1 &    50.0 &    26.0 &    14.0 &   
	518.3 &   228.2 &   118.9 &    60.1 \\
\hline
RF 1.2&
	8.2 &     3.9 &     2.4 &     1.7 &    
	36.2 &    16.4 &     8.6 &     5.3 &   
	156.4 &   69.2 &    37.2 &    20.6 \\
RF 1.8&  12.2 &     6.0 &     3.5 &     2.2 &    
	60.7 &    26.8 &    13.5 &     7.8 &   
	187.1 &   113.5 &    58.6 &    33.9 \\
RF 2.0&  14.2 &     6.7 &     3.8 &     2.4 &    
	69.4 &    29.7 &    15.6 &     9.0 &   
	311.7 &   128.6 &    65.4 &    37.8 \\
RF 2.5&   19.2 &     9.3 &     4.9 &     2.9 &   
	104.0 &    40.8 &    21.7 &    11.4 &   
	452.4 &   201.4 &   106.7 &    59.7 \\
\hline
PME &
	18.5 &     9.9 &     6.1 &     4.9 &    
	81.7 &    40.8 &    24.7 &    17.8 &   
	334.6 &   179.3 &   106.2 &    67.8 \\
\hline
\multicolumn{13}{c}{ }\\*[0.02cm]
\hline
\multicolumn{13}{|c|}{\textbf{Pentium-4 3.4 GhZ (Infiniband)}}\\
\hline
%&\multicolumn{4}{c|}{$8\times 8$ lipids}
%&\multicolumn{4}{c|}{$16\times 16$ lipids}
%&\multicolumn{4}{c|}{$32\times 32$ lipids} \\
nodes & 
\multicolumn{1}{c}{1} & \multicolumn{1}{c}{2} &
         \multicolumn{1}{c}{4} & \multicolumn{1}{c|}{8} & 
\multicolumn{1}{c}{1} & \multicolumn{1}{c}{2} &
         \multicolumn{1}{c}{4} & \multicolumn{1}{c|}{8} & 
\multicolumn{1}{c}{1} & \multicolumn{1}{c}{2} &
         \multicolumn{1}{c}{4} & \multicolumn{1}{c|}{8}
\\
\hline
1.8&    17.4 &     8.2 &     4.3 &     2.4 &    
	76.1 &    33.1 &    16.8 &     8.9 &   
	309.4 &   135.8 &    68.6 &    35.6 \\
2.0&   19.4 &     9.3 &     4.9 &     2.6 &    
	85.7 &    36.8 &    18.8 &    10.1 &   
	349.4 &   152.4 &    77.6 &    40.1 \\
2.5&   27.1 &    12.9 &     6.5 &     3.5 &   
	121.1 &    51.4 &    26.4 &    13.8 &   
	339.0 &   215.6 &   110.4 &    56.5 \\
\hline
SH 1.8&  24.7 &    11.8 &     6.1 &     3.2 &   
	108.8 &    47.1 &    24.0 &    12.5 &   
	398.3 &   192.4 &    97.4 &    49.9  \\
SH 2.0&   27.9 &    13.3 &     6.8 &     3.6 &   
	122.2 &    53.2 &    26.8 &    14.0 &   
	334.3 &   218.1 &   109.9 &    56.5 \\
SH 2.5&   38.8 &    18.6 &     9.6 &     4.9 &   
	169.9 &    73.9 &    37.4 &    19.6 &   
	139.3 &   303.9 &   153.9 &    79.2 \\
\hline
RF 1.2& 11.5 &     5.6 &     3.1 &     1.8 &    
	50.0 &    22.1 &    11.4 &     6.1 &   
	205.4 &    90.3 &    46.4 &    23.9 \\
RF 1.8& 19.2 &     9.0 &     4.7 &     2.6 &    
	83.8 &    36.5 &    18.5 &     9.7 &   
	342.6 &   149.9 &    76.2 &    39.2 \\
RF 2.0&  21.5 &    10.4 &     5.3 &     2.8 &    
	94.7 &    41.1 &    21.1 &    11.1 &   
	388.8 &   169.9 &    86.5 &    44.7 \\
RF 2.5& 30.0 &    14.4 &     7.4 &     3.9 &   
	134.3 &    57.2 &    29.0 &    15.1 &   
	282.8 &   237.5 &   121.2 &    64.2 \\
\hline
PME&   21.8 &    11.0 &     6.1 &     3.9 &    
	96.4 &    45.0 &    24.4 &    14.7 &   
	386.2 &   182.2 &    99.3 &    58.2 \\
\hline
\end{tabular}}
\caption{Hours of wall clock time needed per $1~\mathrm{ns}$ of trajectory.
The different columns correspond to different membrane sizes and different
numbers of nodes (for parallel runs).
The
specified membrane size is the number of lipids per single leaflet. The studied bilayers 
thus contained
twice that number of lipids.}
\label{tabSpeed2}
\end{table}

\begin{figure}
\centering
%\begin{picture}(0,0)
%\put(45,99){\textbf{(a)}}
%\end{picture}
%\includegraphics[width=0.6\textwidth]{fig-1.eps}\\
%\begin{picture}(0,0)
%\put(41,99){\textbf{(b)}}
%%\end{picture}
%\includegraphics[width=0.6\textwidth]{fig-2.eps}\\
%\begin{picture}(0,0)
%\put(45,99){\textbf{(c)}}
%\end{picture}
%\includegraphics[width=0.6\textwidth]{fig-3.eps}
\includegraphics[width=\textwidth]{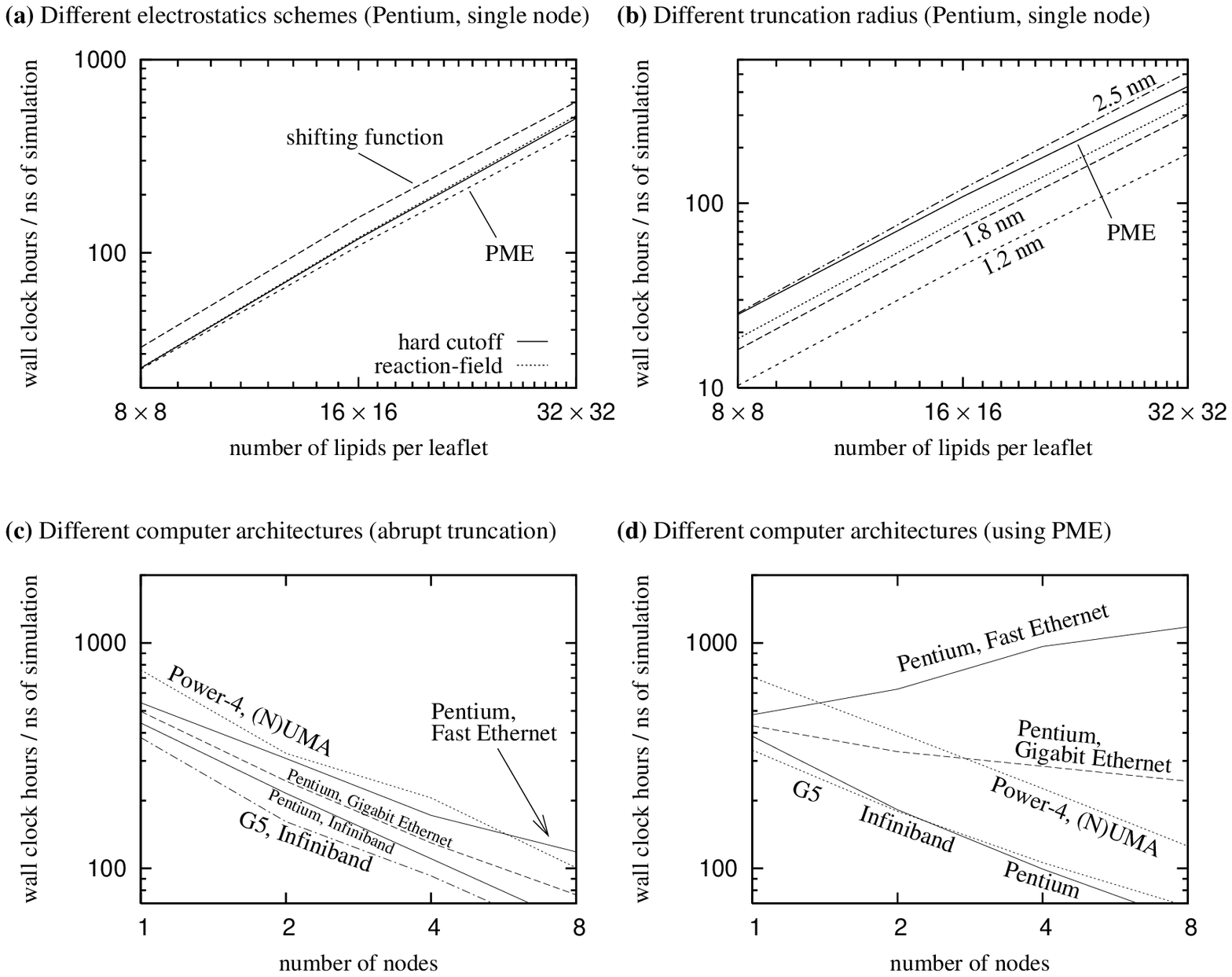}
\caption{\textbf{(a)} The time needed (hours of wall clock time) to compute 
a $1~\mathrm{ns}$ trajectory for membranes of different size. The results are for a 
single Pentium-4 at 2.66 Ghz and $\rcut=2.5~\mathrm{nm}$. 
\textbf{(b)} Comparison of reaction field with different values of $\rcut$.
(Remember that $\rcut=1.2~\mathrm{nm}$ uses a smaller $r_{\mathrm{list}}$.)
\textbf{(c)} Benchmark results for truncation on the different architectures with a varying number of nodes used for parallel runs.
N(UMA) stands for nearly uniform memory access.
\textbf{(d)} Benchmark results for PME runs on the different architectures with
a varying number of nodes used for parallel runs.
Each leaflet contains $16\times 16$ lipids.}
\label{figBenchmark1}
\end{figure}

Simulations of biomolecular systems are typically limited by the lack
of available computer power. 
Tables~\ref{tabSpeed1} and \ref{tabSpeed2} list our benchmark results for different 
architectures and networks. We have studied lipid
systems of various sizes on a varying number of nodes for parallel runs using all
electrostatics schemes described in this paper. The three architectures --
Pentium-4, IBM Power 4 and Apple/IBM G5 --
represent the majority of computer systems used around
the world. We did this with the intent of enabling the user to
estimate the time that his\,/\,her own simulations will need. The user needs
to pick one of our three test systems, closest to his\,/\,her own computational
resources, and then simply scale our benchmark results with the difference in
clock frequency.

First, we tested Pentium-4 CPUs running at 2.4 Ghz, connected by Fast
Ethernet switches. We also used Pentium-4 CPUs running at 2.66 Ghz,
connected by GigaBit Ethernet switches. 
Then, we used the Apple/IBM G5 based systems in which each node 
was a Dual 2.3 GHz PowerPC 970FX processor. The nodes were
connected by Infiniband network. We also tested a network of Infiniband connected
3.4 GHz Pentium-4 CPUs.
Finally, we used Power-4 nodes running
at 1.1 Ghz on an IBM eServer Cluster 1600 supercomputer. The latter has a
nearly uniform memory architecture, i.e., communication between CPUs is
possible by directly reading from or writing to other CPUs' local memory.
% but this
%operation takes longer than operations on own local memory. 
For the eServer
Cluster architecture, the speed difference is only a factor of $2.7$ on the
average (less than the speed difference between different levels of cache on a
single Pentium-4 CPU) such that the name ``almost-uniform memory architecture''
captures the concept well.

\subsection{Single processor performance}

A selection of the single-processor results
from Tables~\ref{tabSpeed1} and \ref{tabSpeed2} (for $\rcut=2.5~\mathrm{nm}$) is shown
graphically in Fig.~\ref{figBenchmark1}a-b. Reaction field is slightly slower than
abrupt truncation, while shifting function needs approximately $30\,\%$ more
time. In contrast to common belief, PME is faster than any of the truncation
methods. We can only offer better cache utilisation as an explanation.
We also found the relative positions of the curves to be almost independent
of computer architecture. For $\rcut=2.0~\mathrm{nm}$, the relative performance
of PME drops slightly and becomes equivalent to the use of shifting functions.
Only for $\rcut=1.8~\mathrm{nm}$ is PME slower than all the 
truncation methods we tested.

In order to understand these results we should recall that we have employed a
twin-range setup in evaluating electrostatic interactions. Within a certain
distance $r_{\mathrm{list}}$, mutual electrostatic interactions (called
short-range interactions) are evaluated at every integration step, while outside
of $r_{\mathrm{list}}$ (called long-range part) this is done only every tenth
integration step. For $\rcut=1.8~\mathrm{nm}$, for example, more than two
thirds of the force evaluations are for short-range interactions. The scheme
for treating the long-range part of electrostatic interactions thus has less
influence than might be expected naively.

For the short-range part $r\le r_{\mathrm{list}}$, a potential with abrupt
truncation is the easiest to compute since $\rcut>r_{\mathrm{list}}$ and thus
only the plain Coulomb's law (without a cutoff) needs to be evaluated. The
reaction field expression~(\ref{eqReactionField}) is more complicated than 
Coulomb's law~(\ref{eqPotential1}). Since the additional terms such as $r^2$ need
to be computed anyhow in order to compute $1/r$, the overhead is 
surprisingly small: the number of floating-point operations increases by less
than $20\,\%$. Since also the Lennard--Jones and bonded interactions and
memory accesses take some time, the observed speed decrease is significantly
lower than this.

The shifting scheme is particular in that the functional form of the potential
changes at $r_{\mathrm{switch}}$. A conditional statement is very expensive due
to the high cost of flushing the CPU pipelines in case of a misprediction. For
this reason, it is more efficient to implement shifting functions via a table
containing $\mathcal{V}_{ij}(r)$ for discrete values of $r$. The lower speed of
shifting functions compared to the other two truncation schemes reflects the
higher computational cost of a table lookup and subsequent interpolation
compared to direct evaluation of a straight-forward potential. 

During the short-range electrostatics evaluation in the switching scheme, 
$\mathcal{V}(r)$ is computed for values of $r$ that can be slightly larger 
than $r_{\mathrm{list}}$. This is because
the test $r<r_{\mathrm{list}}$ is applied to the entire charge group,
and the test is done only during a neighbour-list rebuild. Still, it should be
noted that for $r_{\mathrm{list}}\gg r_{\mathrm{switch}}$ the short-range
electrostatics evaluation can use the computationally cheaper
direct evaluation of the Coulomb potential~(\ref{eqPotential1}). Unfortunately,
specialised treatment for this case is not implemented into GROMACS, and it would
also not be beneficial for us as we use
$r_{\mathrm{switch}}=r_{\mathrm{list}}$.

When PME is employed, the short-range part of the potential is given by the
complementary error function. Since this function is very expensive to compute, a tabulated
potential is used. In view of the speed of tabulated shifting functions, the
very competitive speed of PME demonstrates that it is a very efficient method
for computing the long-range part of the potential -- its speed is able to
compensate for the slowness of the short-range evaluation.

As discussed above, more time is spent on computing the short-range part 
of the potential
than on the long-range part. The total number of electrostatic
evaluations thus only increases by a factor $1.6$ in going from
$\rcut=1.8~\mathrm{nm}$ to $\rcut=2.5~\mathrm{nm}$ instead of the naively
expected factor $(2.5/1.8)^3\approx 2.7$. An example for the time increase
with increasing $\rcut$ is shown in
Fig.~\ref{figBenchmark1}b for reaction field. (Remember that for
$\rcut=1.2~\mathrm{nm}$ we employ $r_{\mathrm{list}}=0.9~\mathrm{nm}$ instead of
the standard value $r_{\mathrm{list}}=1.0~\mathrm{nm}$. Most of the speed
increase seen in the figure for $\rcut=1.2~\mathrm{nm}$ stems from this.)

\subsection{Scalability}

Apart from the intrinsic speed of an electrostatics scheme, its
parallelisability is of great importance. Over the past years, the size of
the biosystems that are studied has increased faster than the speed of a single CPU,
thereby  increasing the need for parallel simulations.
Figure~\ref{figBenchmark1}a-c show a few selected benchmark results for PME and
abrupt truncation. The scaling results for all truncation schemes are basically
identical, and scaling is influenced surprisingly little by the size of the
lipid membrane. A closer analysis of Tables~\ref{tabSpeed1} and \ref{tabSpeed2} shows that on the
IBM eServer supercomputer, especially for reaction field, quite a few instances of
superlinear scaling are observed.  The appearance of superlinear scaling itself
can be explained by an increase of the total available cache space with
increasing number of nodes but we cannot offer an explanation why this is more
prominent for reaction field than for the other truncation methods.
Figure~\ref{figBenchmark1}d shows that truncation methods scale 
very well on all the architectures we studied. On a supercomputer, 
PME scales as well as the truncation methods. However when PME was used on
Gigabit Ethernet, the wall clock time needed decreased only slightly when 
the number of nodes is increased, and with Fast Ethernet the time actually
increased with number of nodes. 
These results for different network architectures let us to conclude that
using GROMACS (at least), it is thus senseless to use parallel 
PME if only Fast Ethernet is available and it is doubtful even when using 
Gigabit Ethernet.

This can partly explain why relatively little attention has been 
given to benchmarking parallel electrostatics until recently. 
In the days when most numerical work was done at a few supercomputing 
centres, every electrostatics scheme would perform approximately 
equally well. Only with the advent of cheap Linux clusters, 
the problem of poor scaling of, e.g., PME arose.

%It is important to keep in mind that, strictly speaking, the poor 
%scaling of PME found here is a property of the the present version 
%(version 3.2) of
%GROMACS. Due to the lack of similar data for other simulation 
%packages such as NAMD or Amber, we are unable to conclude to what extent 
%this poor scaling is due to PME and how much it is related to the 
%decomposition scheme used in the parallelization of the program.
%, 
%as well as the implementation of PME for parallel simulations. 
%Nevertheless, care is warranted when PME is used in parallel simulations
%in GigaBit and FastEthernet networks. 
%%
The numerical work for computing the long-range part of the 
potential is mainly given by the work needed for the FFT. As all
commonly used programs employ the same FFTW library, they have in principle
the same limitations for scalability. Programs differ, however, as well as
different methods like PME vs. P3M differ, in their distribution of work 
between the Fourier part and the direct summation. The distribution employed by
GROMACS, which by default does not depend on the number of nodes, is somewhat
suboptimal in this aspect. Other programs
might improve scalability by
shifting more work to the direct summation while increasing the FFT grid
spacing, thereby reducing the work done by the FFT that is parallelising worse
 than the direct summation.

\section{Discussion}
\label{secDiscussion}

In this paper, we have used a lipid membrane as test bed to examine various
schemes commonly used to handle electrostatics  in atomistic MD simulations of
biomolecular systems.  We compared different truncation schemes (abrupt
truncation, shifting function and reaction field) and the particle-mesh Ewald
technique.
The focus here was on speed and performance on different architectures and networks.
%
%, both with respect to their effect on the properties of the bilayer
%as well as with respect to their speed on different hardware. We should stress
%that 
All of these schemes have been treated on an equal footing, i.e., 
all interactions have been identical except for the scheme used for
computing the electrostatic interactions. 

On the hardware side, the Apple/IBM G5 based system performed very well showing
the strengths of the architecture. Earlier this year, Apple started moving
from the IBM G4/G5 chips to Intel processors which seems to mark the 
end of an era. Similarly, another good performer, the DEC Alpha processor,
was phased out a few years back.

On the physical side,  an abrupt truncation
of electrostatic interactions should be avoided~\cite{Patra:03vl,Anezo:03mz,Patra:04te}. It introduces major artifacts
into the system without offering any noteworthy benefits in speed compared to,
e.g., reaction field techniques.
Shifting functions, reaction field technique and PME all offer results of good
quality even though the quantitative results disagree somewhat.
This can be fixed, e.g., by tuning the force field but care should be taken:
recent studies have shown that even the RF method should be used with great
care in cases of bulk water\,\cite{Yonetani:06xo}.
Of
more interest is thus the following question: given a force field developed
for some particular electrostatics scheme, which other schemes can be
substituted without the need to reparameterise the force field?

We also found (data not shown) that the results for the shifting function and reaction field are
only moderately dependent on $\rcut$. (The only exception to this statement is
$\rcut=1.2~\mathrm{nm}$. In our opinion, $\rcut=1.2~\mathrm{nm}$ should
be avoided in membrane simulations altogether.)
This dependence seems to be even a bit smaller for the reaction field than for
the shifting function. This, however, depends on the weight assigned to the different
quantities studied in this paper. This weak dependence on $\rcut$ allows for
choosing a small value of $\rcut$. On the other hand, our
benchmarks demonstrate that the cost of using a larger cutoff is smaller than
might be naively expected. 

From studying a large number of different quantities that could not be
displayed here for space restrictions, we found that RF and PME are interchangeable,
i.e., going from one scheme to the other changes the results only by a margin
that is comparable with experimental uncertainty. This property is the main
advantage of reaction field over shifting functions. There is, however, one
important question:
given a force field developed for some particular
electrostatics scheme, with a particular value of the $r_c$ for 
a truncation method,
is it possible to change that value or even to choose a
completely different scheme -- without the need to reparameterise the force
field? Here, we have not tried to answer that question, but it should
be kept in mind whenever different methods with various force fields
are being chosen and tested.

%The difference between
%reaction field and shifting function comes somewhat unexpected as the force (cf.
%Fig.~\ref{figPotential}) is almost identical.

%Here, we have focused on the performance of Ewald summation based approach
%vs the use of direct cutoff and reaction field. 
Our testbench in this study was the
GROMACS simulation engine. As mentioned in the beginning, GROMACS is by 
no means the only possible simulation engine. However, we expect the trends be
largely
similar independent of the software package. In the case of Ewald summation
based electrostatics, performance depends heavily on the Fast Fourier Transform
library. 
The FFTW library~~\cite{Frigo:05st} is a common choice, and packages such
as GROMACS, Amber~\cite{Case:05mc}, NAMD~\cite{Kale:99ia,Phillips:05uq}, LAMMPS~\cite{Plimpton:95uc}, and  Espresso~\cite{Limbach:06xw} use FFTW.
Some of the packages also offer other options for performing FFTs, but the
FFTW library is by far the most popular in the field.

There are also other methods for evaluating electrostatic interactions, but these are 
not as widely used as PME. The P$^3$M method is 
another efficient Ewald-based scheme. Other developments on the Ewald-based
methods have appeared recently~\cite{Nam:05od,Shan:05td}. 
Of the methods that are not based on Ewald summation,
the Fast Multipole Method,~\cite{Greengard:97vg,Hrycak:98xw,Ying:04uy} often called just FMM, is 
a different set of methods utilising the multipole expansion. 
It offers better scaling, as it is an $\mathcal{O}(N)$ algorithm. This, however,
comes with the expense of larger prefactors. Multigrid 
methods~\cite{Sagui:01ia,Izaguirre:05qe} are
also newcomers, offering, like FMM, $\mathcal{O}(N)$ scaling. These methods 
also natively include other boundary conditions than the usual periodic
ones. Another novel $\mathcal{O}(N)$ algorithm is 
the method of Maggs \textit{et al.}~\cite{Maggs:02fi,Rottler:04jw} which
is based on propagating the electric field and solving the Maxwell's equations.
However, we did not test the above non-Ewald algorithms as standardised (\'a la
GROMACS or NAMD)
implementations do not (to our knowledge) exist.

Finally, it is worth noticing that Ewald summation -based
methods are typically fast and suitable only in fully periodic systems.
If periodicity is broken, e.g., by the presence of a wall, the method
needs to be modified. This strongly affects the performance. Alternatively, 
other methods need to be used. For Ewald-based methods there
are several possible variants. These typically use the trick of adding an extra
dipolar term, and including empty space in the simulation box~\cite{Yeh:99zz}.
A review and comparison of many of the earlier Ewald-based methods 
for slab geometries has been written by Widmann and Adolf~\cite{Widmann:97wq}. 
There have been many new developments to include boundary conditions
other than periodic. Of the above methods, FMM, multigrid, and 
the methods of Maggs \textit{et al.} natively include other boundary conditions. Other 
methods include Lekner summation~\cite{Lekner:91op} and the MMM2D method~\cite{Arnold:02gc}.

In summary, long-range interactions, such as the Coulomb interaction,
still pose challenges to simulators. This is clearly demonstrated by the large number 
of articles published on these topics over the last few years. Modellers need to take care in choosing
an appropriate  method, one that maximises computational efficiency, at the same time
giving reliable results.
 
%Here, we have studied
%scalability of different commonly used electrostatic summation methods
%using different computer architectures and networks using the GROMACS simulations suite as the testbed. 
%The three architectures Ð Pentium-4, IBM Power 4 
%and Apple/IBM G5 Ð represent the majority of computer systems used around 
%the world. We did this with the intent of enabling the reader to estimate the 
%time that his/her own simulations will need. He/she needs to pick one of our 
%three test systems, closest to his/her own computational resources, and then 
%simply scale our benchmark results with the difference in clock frequency.
\subsection*{Acknowledgements}
%\mbox{}\hspace{2cm}\textbf{Acknowledgments}

This work has been supported by the Academy 
of Finland Centre of Excellence Program, 
%(E.\,F. and I.\,V.), 
%the National Graduate School in 
%Materials Physics (E.\,F.), 
the Academy of Finland, 
%Grants No.~00119, 54113 (M.K.) and No.~80246 (I.V.), 
the 
Jenny and Antti Wihuri Foundation, 
%(M.H.), 
the Federation 
of Finnish Insurance Companies, 
%(M.H.), 
the Emil Aaltonen Foundation,
% (M.K.),
the European 
Union Marie Curie fellowship HPMF--CT--2002--01794 
(M.\,P.), European Program MRTN-CT-2004-512331 (M.\,P.),
and by the Beckman Institute Fellows program (E.\,F.).
 We would also like to thank the Finnish IT Centre 
for Science (CSC), the supercluster computing facility at 
the University of Southern Denmark, Center for Parallel Computers at the
Royal Institute of Technology in Stockholm, 
the Terascale Computing Facility at Virginia Polytechnic
Institute \& State University for computational resources,
and the Laboratory for Computational Engineering at the
Helsinki University of Technology.

%\bibliographystyle{elsart-num}
%\bibliography{/Users/mikko/Documents/Articles/Bib/biblio-mk-new}
%\bibliography{biblio-mk-new}

\end{document}